# Magnetoresistance of monolayer graphene with short-range scattering

G. Yu. Vasileva[1,2,3], P. S. Alekseev[1], Yu. B. Vasilyev[1], Yu. L. Ivanov[1], D. Smirnov[2], H. Schmidt[2], R.J. Haug[2]

[1] *Ioffe Institute, St. Petersburg, 194021 Russia*

[2] *Institut für Festkörperphysik, Universität Hannover, Appelstraße 2, 30167 Hannover, Germany*

[3] *Peter the Great Saint-Petersburg Polytechnic University, St. Petersburg, 195251 Russia*

We present magnetotransport measurements at classical magnetic fields for three graphene monolayers with various levels of disorder. A square root magnetoresistance (SRMR) behavior is observed in one sample which has the characteristic sub-linear conductivity signaling on the presence of short-range disorder in this sample. No square root MR was observed in other samples where short-range scattering is inessential as it is evident from the gate voltage dependences of their conductivities. Comparing our experimental data for the sample with theoretical calculations we found a good qualitative agreement and established the conditions which should be fulfilled in graphene to observe the SRMR experimentally.

It is believed that the transport properties of exfoliated graphene placed on a $SiO_2$ substrate are mainly defined by carrier scattering on charged impurities [1,2]. Indeed, usually the conductivity $\sigma$ is linear versus the back gate voltage $V_g$. This dependence is ascribed to electron scattering on impurities with a long-range Coulomb potential [1,2]. However, in some samples $\sigma$ is observed to be sub-linear in $V_g$ that is explained by the contribution of short-range scattering (resulting, for example, from point defects) [3,4]. Recently, it was predicted [5] that for the case of short-range scattering the magnetoresistance (MR) has interesting characteristic features owing to the unusual linear energy dispersion in monolayer graphene. In particular, a square root MR (SRMR) was revealed both at very low and at very high magnetic fields $H$: $[\rho_{xx}(H)-\rho_{xx}(0)]/\rho_{xx}(0) \approx C\sqrt{H}$. Such the square root dependence was observed experimentally [6] at low magnetic fields but away from the Dirac point whereas the theory [5] implies that the chemical potential lies near the Dirac point. A detailed comparison between experiment and theory is necessary for a further understanding of this effect.

The emergence of the square-root magnetoresistance at weak magnetic fields is a significant effect. Its importance is determined by the fact that this is a new type of magnetoresistance which has not been experimentally studied before. Such a magnetoresistance can be realized only in systems with a linear dispersion, and it is absent in conventional semiconductors with a parabolic dispersion, where the square-root magnetoresistance occurs only at strong fields (in the quantum regime) [7-9].

In this paper we present magnetotransport data for three graphene monolayer samples with qualitatively different behavior of their conductivity with gate voltage. We found a square root MR behavior in one sample (sl1) which has the characteristic sub-linear conductivity signaling on the presence of short-range disorder in this sample and make detailed analysis of the SRMR data using the theory of Ref. [5]. No square root MR was observed in other samples (sl2, sl3) where short-range scattering is inessential as it is evident from the dependences of their conductivities on $V_g$. Comparing our experimental data for the sample sl1 with theoretical calculations we found a good qualitative agreement in the full range of magnetic fields and established the conditions which should be fulfilled to observe the SRMR experimentally.

The graphene monolayers were fabricated by micromechanical cleavage of natural graphite. We used Si with a 330nm silicon oxide wafer as substrate. Several monolayer graphene samples were investigated. Monolayers were identified by their optical contrast and confirmed afterwards by measurements in high magnetic fields, where Shubnikov-de Haas oscillations show distinct minima and the Hall voltages have plateaus at filling factors 2, 6, 10, 14 indicating that our graphene sample is indeed a monolayer as shown for sample sl2 in Fig.1b. One of the reported samples (sl1) was annealed at 300°C in $He_2$ atmosphere during 3 hours. Another sample (sl2) was cleaned by an AFM tip that allows to decrease considerably the concentration of the charged impurities on top of graphene. The third sample (sl3) was measured without annealing, as prepared. The carrier mobility for sl1 and sl2 was estimated as 6500 $cm^2$/ Vs and for sl3 as 1000 $cm^2$/ Vs. Two samples (sl2 and sl3) were prepared in the Hall bar geometry and sample sl1 has an irregular shape. The carrier concentration $n$ in the samples can be varied up to $5\times10^{12}$ $cm^{-2}$ by applying voltage to the conducting substrate, which acts as a back gate. The magnetoresistance was measured by lock-in amplifiers with low-frequency currents (f = 17.7 Hz, I = 100 nA) in a temperature range 1.5-200 K and at magnetic fields up to 12T.

Figure 1a demonstrates the gate-voltage dependences of the conductance $\sigma(V_g)$ for samples sl1 (red line), sl2 (blue line) and sl3 (green line). The samples have quite different chemical doping which is notable from the position of the Dirac point ($V_D$) in Fig.1a: The sample sl3 has $V_D$ of 38V which is far away from $V_D$ of 5V and 8V in samples sl1 and sl2, respectively. The dashed straight lines in Fig. 1a are drawn to demonstrate that the conductance $\sigma(V_g)$ is completely different for all three samples: For sl1 conductance varies sub-linearly, it is super-linear in sl2, and it is proportional to $V_g$ for sl3 throughout the full range.

The difference of the conductance behaviour in Fig. 1a reflects the difference of scattering mechanisms in these samples [1-4]. Analyzing the conductance behaviour in Fig. 1a one can estimate the significance of different mechanisms of carrier scattering. A linear and super-linear behaviour of $\sigma(V_g)$ corresponds to the regime when the particle scattering on charged impurities is the dominant long-range scattering mechanism [10]. We conclude that the long-range Coulomb impurities are dominated in samples sl2 and sl3. It was also experimentally found that graphene samples can exhibit sub-linear behavior of their conductivity as a function of $V_g$ [3,4] and that sub-linearity of $\sigma(V_g)$ is an inherent property of high-mobility samples [4]. In particular it was shown [11, 12] that the degree of sub-linearity of $\sigma(V_g)$ increases when the contribution of short-range defects increases comparatively to the Coulomb scattering. Furthermore, as indicated in Ref. [13] the asymmetry in the conductivity curves as it is observed in sample sl1 (see red curve in Fig. 1a) also points on the importance of short range defects. So the strong sub-linear behaviour of $\sigma(V_g)$ at gate voltages below 2 V and the asymmetry above 10 V in the conductivity curve in sample sl1 give evidences that scattering on short-range defects is more significant in the sample sl1 than in the samples sl2 and sl3.

Detailed experimental resistance measurements at different magnetic fields and gate voltages were done for all samples. The results are shown for samples sl1, sl2 and sl3 in Fig. 2, Fig. 3 and Fig. 4, respectively. No SRMR was found in samples sl2 and sl3. It is not surprising as far as in these samples signs of short-range scattering were not seen. To demonstrate this, we selected several curves of magnetoresistance both for "electron" and "hole" sides of conductivity curve not far away from the Dirac point (Fig. 3b). Signatures of SRMR are not seen but only linear and super-linear behaviour of MR. For the sample with a high level of defects sl3 (Fig. 4b), negative magnetoresistance always dominated with the exception of some range of gate voltages corresponding to high hole densities at temperatures T>150 K. At the same time it is clearly visible that the SRMR at low magnetic fields (below 20 kOe) exists only in sample sl1 in the "hole" region of conductivity curve in a narrow range of $V_g$ between 0V and -2V (Fig. 2b). The data of the MR (dashed lines) proportional to $H^{1/2}$ are presented in Fig. 2b together with SRMR fitting (solid lines). Outside these $V_g$ values we do not observe SRMR. It is important that, in particular, no SRMR was found for the "electron" region of the conductivity at all studied $V_g$ values in this sample. The difference between "electron" and "hole" branches of conductivity in sample sl1 (Fig. 1a) where the mobility of electrons is considerably lower than for holes (at gate voltages larger than 10 V) may be responsible for this fact. Using our experimental results we can list conditions which should be fulfilled to observe the low magnetic field SRMR. First, SRMR is realized only in samples with essential short-range scattering so that it makes the conductivity to be sub-linear as a function of gate voltage. Second, it is observed away from the Dirac point because at the Dirac point the linear MR is realized due to equal numbers of electrons and holes as it is shown in Ref. 14. At the same time far away from the Dirac point the effect of SRMR is very weak in accordance with the exponential decrease with the Fermi energy as it was shown in Ref. 5. So SRMR may be observable only in a narrow range of gate voltage.

We would like to comment why in experiments the SRMR is realized away from the Dirac point. In addition to the recent theory [14] explaining the linear MR at the Dirac point, we can propose two possible reasons. First, as it was discussed above, in real samples carrier scattering on the long-range Coulomb impurities plays an important role. The rate of scattering on the long-range Coulomb impurities increases with decreasing the carrier energy and the rate of scattering on the short range defects decreases with decreasing of the carrier energy, according to the formulas [1]: $\frac{\hbar}{\tau_{short}(\varepsilon)} \propto \varepsilon$, $\frac{\hbar}{\tau_{Col}(\varepsilon)} \propto \frac{1}{\varepsilon}$. Due to this fact, the influence of a short range potential in the Dirac point should be minimal, and hence the basic condition for the appearance of the square-root dependence is hardly satisfied. Second, another assumption of the theory [5] is that the temperature (or the "effective temperature" discussed above) is much higher than the separation between neighbouring Landau levels near the Fermi energy. As the energy shifts away from the Dirac point, the Landau levels in graphene become closer to each other owing to their uneven spacing $E_n \sim n^{1/2}$. This condition corresponds to the situation when the chemical potential lies away from the Dirac point.

The rest of the paper is devoted to a detailed analysis of the data for the sample sl1 with the short-rang scattering. The dependence of the magnetoresistance for the sample sl1 at $V_g = 0$ is plotted in Fig. 5 throughout the total range of magnetic fields. The minimum of MR around $H=64$ kOe is related to Shubnikov-de Haas oscillations and will not be discussed. Three parts of the plot with different behavior are clearly visible in the region of positive magnetoresistance. In addition to a square-root dependence $\rho(H)/\rho_0 \sim H^{1/2}$ observable at weak (up to $H=16$ kOe), and high (from $H=30$ kOe) [15] magnetic fields there is an interval between them where the magnetic field dependence of the resistance changes and has negative curvature (between 16 and 30 kOe). We fitted $\rho(H)/\rho_0 \sim H^2$ in Fig. 5 for this range however the fitting is ambiguous and $\rho(H)/\rho_0 \sim H^{3/2}$ also fits the experimental data quite well (not shown). It is impossible to indentify which of these two cases is realized experimentally because the inflection points separating regions with convexity and concavity lie close to each other.

Let us compare the obtained experimental data on monolayer graphene magnetoresistance with the theory [5] developed for short-range scattering. The key energy scale is given by the formula:

$$\varepsilon^* = \sqrt{\gamma} \frac{\hbar \text{v}}{l_m} = \sqrt{\gamma} \frac{\sqrt{\hbar e} \text{v}}{\sqrt{c}} \sqrt{H} \quad (1),$$

where v is the electron velocity of carriers in graphene, $l_m$ is the magnetic length, $\gamma = \varepsilon \tau(\varepsilon)/\hbar$ is the dimensionless parameter of the theory, which does not depend on $\varepsilon$ for scattering on short-range impurities; $\tau(\varepsilon)$ is the quantum lifetime of an electron in a given state. At energies, lower than $\varepsilon^*$, the electron Landau levels are separated, i.e. at the energy $\varepsilon = \varepsilon^*$ there is a transition from classical to quantum character of electron states. This is the particular feature of the Dirac dispersion (as in graphene) where the energy spectrum is strongly nonparabolic and the Landau level separetion drastically decreases with energy. For the data presented in Fig. 5 $\mu \gg T$ ($\mu =$ 66 meV is the chemical potential and $T=$ 4.7 meV is the temperature), the magnetoresistance curve has a square root dependence both at weak magnetic fields $\varepsilon^* < \varepsilon_2 = (\mu^2/T)\exp(-\mu/3T)$ and at high magnetic fields $\varepsilon^* > \mu$. At weak $H$ field MR is given by the formula:

$\Delta\rho/\rho_0 = 3.18\varepsilon_* \exp(-\mu/T)/T$ (2).

Between these two parts MR shows a quadratic behavior $\sim \varepsilon^{*4}$ ($\sim H^2$). Due to the small quantities of $\exp(-\mu/T)$ MR is exponentially small and is expected to be hardly observed experimentally. Nevertheless, as it is seen from Fig. 5, the square-root dependence is visible and the qualitative coincidence between the experimental curve and the theoretical behavior is roughly fulfilled. To explain this, we suppose that instead of the sample temperature $T$ an effective temperature $T_{eff}$, $T_{eff} \gg T$, should be taken into account. We can associate the effective temperature $T_{eff}$ with the disorder potential due to the inhomogeneity of the sample. Roughly speaking, inhomogeneity gives different origins for the cone energy spectrum of graphene in different space points, therefore the Fermi level being constant everywhere in absolute terms, is different if calculated from the bottom of the cone in a given space point. The long-range random potential of large amplitude acts effectively as an increase in the effective temperature $T_{eff}$, which exceeds $T$ in our samples. Existence of such smooth potential in graphene samples was pointed out in Ref. 12.

A rough estimation using the temperature dependence of conductivity [16] defines the effective electron temperature as $T_{eff} = 20$ meV. We fit the weak magnetic part in Fig. 5 by a square root dependence (2). Using the formula (1) for $\varepsilon_*$, we obtain the value of the parameter $\gamma$ of the short-range potential from this fitting: $\gamma = 3.6$. The theory [5] implies that $\gamma \gg 1$. Such the result is quite reasonable, but indicates that the experiments were performed at conditions corresponding the boundary of applicability of the theory. Furthermore, we can calculate the value of $\varepsilon_2$. This energy separates two different regimes where MR is either proportional to $H^2$ or to $H^{1/2}$. The $H^2$ regime is the well known "classical" MR which is related to the dependence of the electron parameters on energy and to temperature smearing of the electron energy distribution. The $H^{1/2}$ regime is defined by electrons at high energies of the Dirac cone that gives $H^{1/2}$ dependence of MR as it was shown in Ref. 5. These two effects in MR become equal at a magnetic field corresponding to $\varepsilon_2$. Using the expression $\varepsilon_2 \approx \frac{\mu^2}{T}e^{-\mu/3T}$ we have $\varepsilon_2 \sim \mu$. Indeed from the parameters of our experiment ($\mu \sim 60$ meV, $T_{eff} \sim 20$ meV) it follows that $\mu \sim 3T_{eff}$, and we directly get $\varepsilon_2 \approx \frac{\mu^2}{2.7T} \sim \frac{\mu^2}{\mu} \sim \mu$. This finding indicates on the important role of the chaotic potential in our samples and proves correctness of utilization of the effective temperature $T_{eff}$ for estimations.

Fitting of other regions of the curve in Fig. 5 by formulas from the theory [5] is ambiguous due to the small width of the regions and poor manifestation of the features but qualitatively the negative and positive curvatures are consistent with the theoretical prediction [5].

It is important to explain the difference in interpretation of the square root MR at high and weak magnetic fields. The square root magnetoresistance at high fields for 2D electrons was predicted for the first time for a parabolic energy spectrum electrons [7,8], and the theory [7] can be applied to graphene samples also with some inessential modifications [9] (see also [5]). In contrast, the square root MR at weak magnetic fields is realized only for particles with a linear dispersion, such as in graphene [5]. This is due to the fact that the effective mass of particles strongly depends on their energy, and the carrier scattering time is inversely proportional to their energy [5].

To conclude, experimental measurements for various graphene monolayers give qualitatively different magnetoresistance behaviours for samples with different dependences of conductivity on $V_g$. This difference is naturaly ascribed to different scattering mechanisms as long-range Coulomb and short-range scatterings. Application of the theory of graphene magnetoresistance developed for short-range scattering to our data gives qualitatively good agreement for the higher-mobility samples away from the Dirac point where short–range scattering plays an essential role. All three different parts of the magnetoresistance curve defined by theory are observed experimentally.

We thank I. V. Gornyi, A. P. Dmitriev, V. Yu. Kachorovskii for fruitful discussions. The work was supported by DFG Priority Programme "Graphene" (Schwerpunktprogramm, SPP 1459), FP7-PEOPLE-2013-IRSES, Russian Foundation for Basic Research (RFBR), programs of the Russian Academy of Science and the "Dynasty" Foundation.

FIGURES

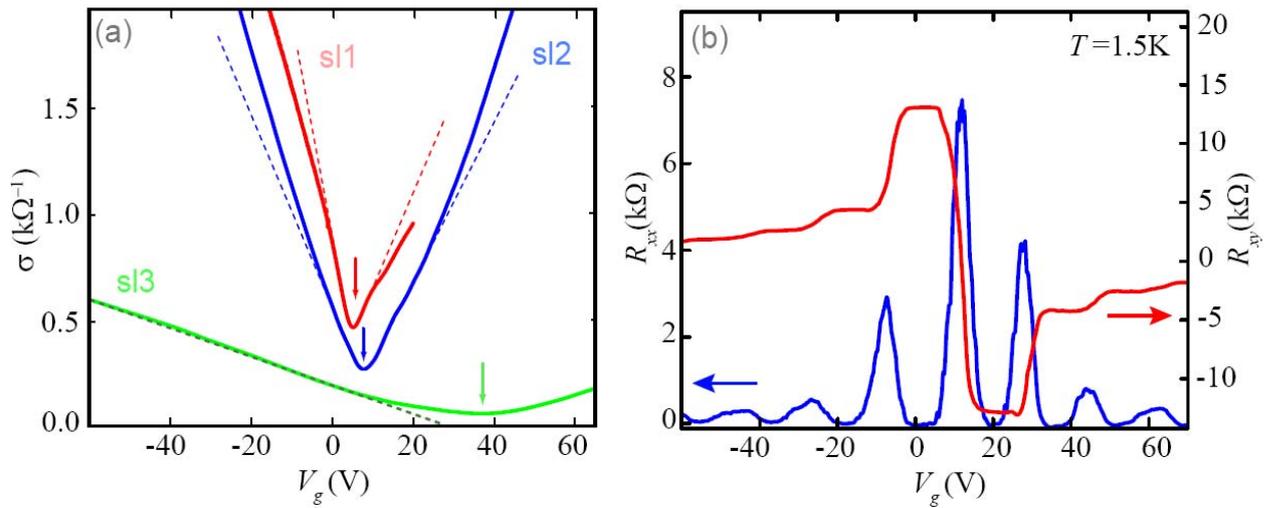

Fig. 1. a) Conductance versus the gate voltage $V_g$ at zero magnetic fields for the sample sl1 (red) sl2 (blue) and sl3 (green). Dashed lines are drawn to demonstrate where the conductance deviates from linear. b) Longitudinal Rxx and Hall resistances measured for the sample sl2 at T=1.5K and H=120 kOe.

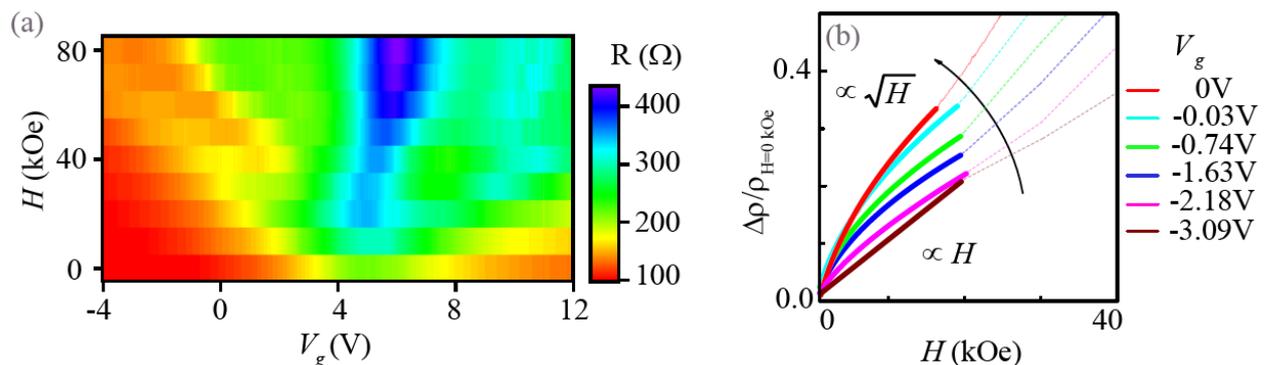

Fig. 2. (a) Magnetoresistance of the sample sl1 plotted as a function of the magnetic field H and gate voltage $V_g$ (b) Several MR curves for different values of $V_g$ together with the best fits (solid lines).

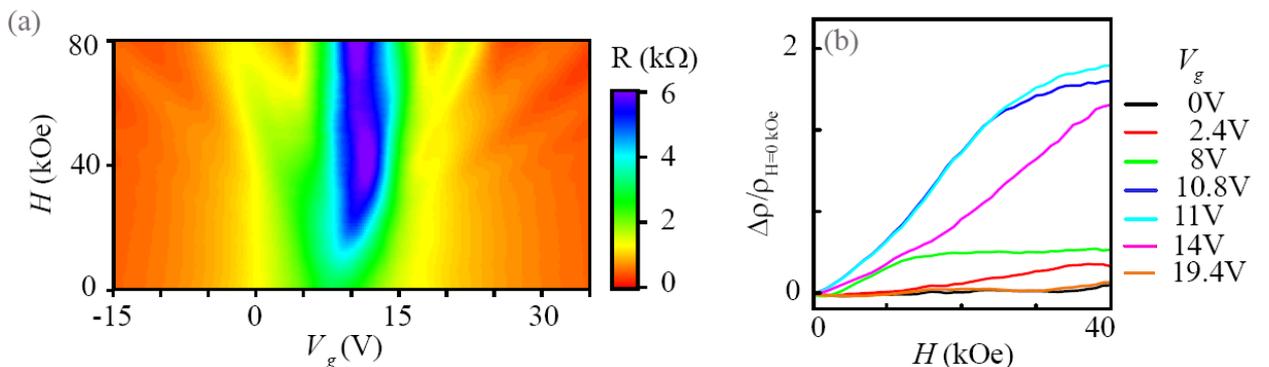

Fig. 3. (a) Magnetoresistance of the sample sl2 plotted as a function of the magnetic field H and gate voltage (b) Several MR curves for different values of $V_g$.

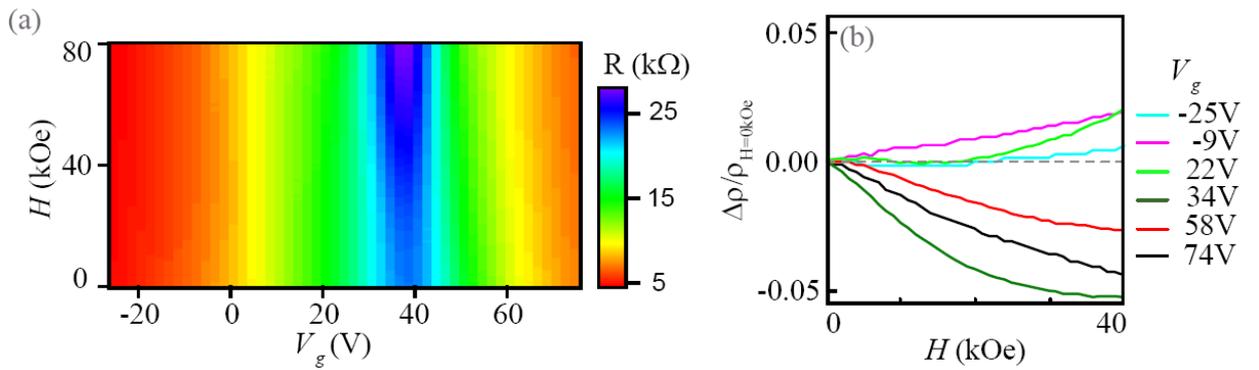

Fig. 4. (a) Magnetoresistance of the sample sl3 plotted as a function of the magnetic field H and gate voltage (b) Several MR curves for different values of $V_g$.

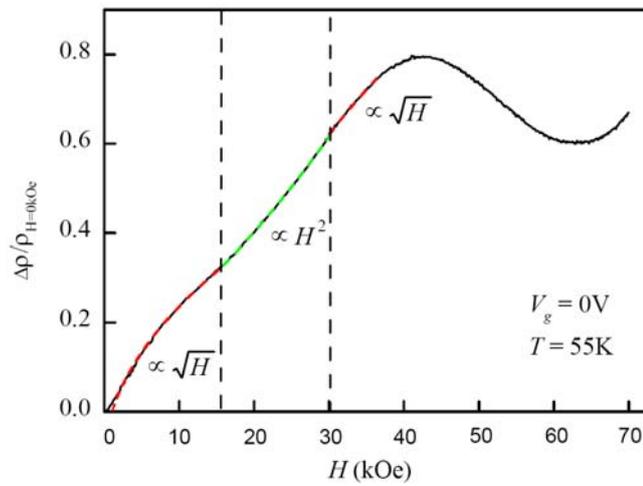

Fig. 5. Magnetoresistance of the sample sl1 (with shot-range scattering) shown in the full range of magnetic fields.